\documentclass[12pt,preprint]{aastex}

\usepackage{amsmath}
\usepackage{amssymb}
\usepackage{graphicx,graphics}
\usepackage[english]{babel}
\usepackage{color}

\def\PGPU{$\varphi$GPU }
\def\PGRAPE{$\varphi$GRAPE }

\begin{document}

\title{Efficient Scheme for Active Particle Selection in N-body Simulations}

\author{
       Shiyan Zhong   \altaffilmark{1}
       }

\altaffiltext{1}{National Astronomical Observatories of China, Chinese Academy of Sciences, 20A Datun Rd., Chaoyang District, 100012, Beijing, China}

\shorttitle{Active Particle Selection}
\shortauthors{Zhong}


\begin{abstract}

We propose an efficient method for active particle selection, working with Hermite Individual Time Steps (HITS) scheme in direct N-body simulation code \PGRAPE. For a simulation with $N$ particles, this method can reduce the computation complexity of active particle selection, from $O(N\cdot N_{step})$ to $O(\overline{N_{act}}\cdot N_{step})$, where $\overline{N_{act}}$ is the average active particle number in every time step which is much less than $N$ and $N_{step}$ is the total time steps integrated during the simulation. Thus can save a lot of time spend on active particle selection part, especially in the case of low $\overline{N_{act}}$.

\end{abstract}


\section{Introduction}

Direct $N$-body code is a useful tool to study dynamics of star clusters and galaxies. The word ``direct" means we compute pairwise forces for all the particles involved in the simulation, which results in $O(N^2)$ computation complexity. Using Hermite Individual Time Steps (HITS) scheme can reduce the computation complexity, because in every time step the code only integrate a small fraction of particles instead of all \citep{Makino1991}. These particles are called ``active particles". Thus at the beginning of every integration loop, one should find out the ID of these active particles.

A simple way to select active particles involves two steps. First evaluate the variable \textbf{min\_t}, which is the minimum value over all $t_i + \Delta t_i$ (we use ``active time" to refer this quantity). Then select those who have $t_i + \Delta t_i$ equals \textbf{min\_t} as active particle. In this method, the same array of data is scanned twice in order to find a few particle IDs. Using the formula given by ~\cite{BNZ2011}, $\overline{N_{act}} \propto N^{0.613}$, we estimate the efficiency (or hit rate) of this method, $\epsilon = \overline{N_{act}} / N \propto N^{-0.387}$. As the total particle number increases, $\epsilon$ drops down and a lot of time is wasted. Doing the selection in parallel can help to reduce the time cost, which has been implemented in \PGRAPE code \citep{Harfst2007}. This three-step selection algorithm works as follows: first, the active time array is divided into $N_p$ equal length segments, where $N_p$ is the number of MPI processes. Each process deal with one segment, find out the local minimum \textbf{min\_t\_loc}. Second, use MPI\_ALLREDUCE function to get the global minimum \textbf{min\_t} and broadcast to all processes. Last, each process do selection work with its own section, then root process collect the information from all processes and broadcast the complete list (another way is to scan the whole array and no more MPI calls are involved). Parallel selection can reduce the computation cost to $O(N/N_p)$. However, in some extreme cases where $\overline{N_{act}}$ is just order of a few, the communication time between different nodes takes over and becomes a bottleneck. Parallel processing can cut down the execution time by a factor of $N_p$ in the best case, however, it does not improve the efficiency.

In this article, we propose another approach to select active particles, as described in next section.

\section{Method}

In the old selecting scheme, data are organized in 1-D array. We introduce a 2-D linked list to organize the particle ID (PID). First we define two kinds of node: T-node and P-node. T-node is used to store the time information and the route to corresponding P-node (a pointer). P-node only store PID. Hereafter, we refer to this new method as Linklist method.

\begin{figure}[htbp]
  \begin{center}
  \includegraphics[width=0.8\columnwidth]{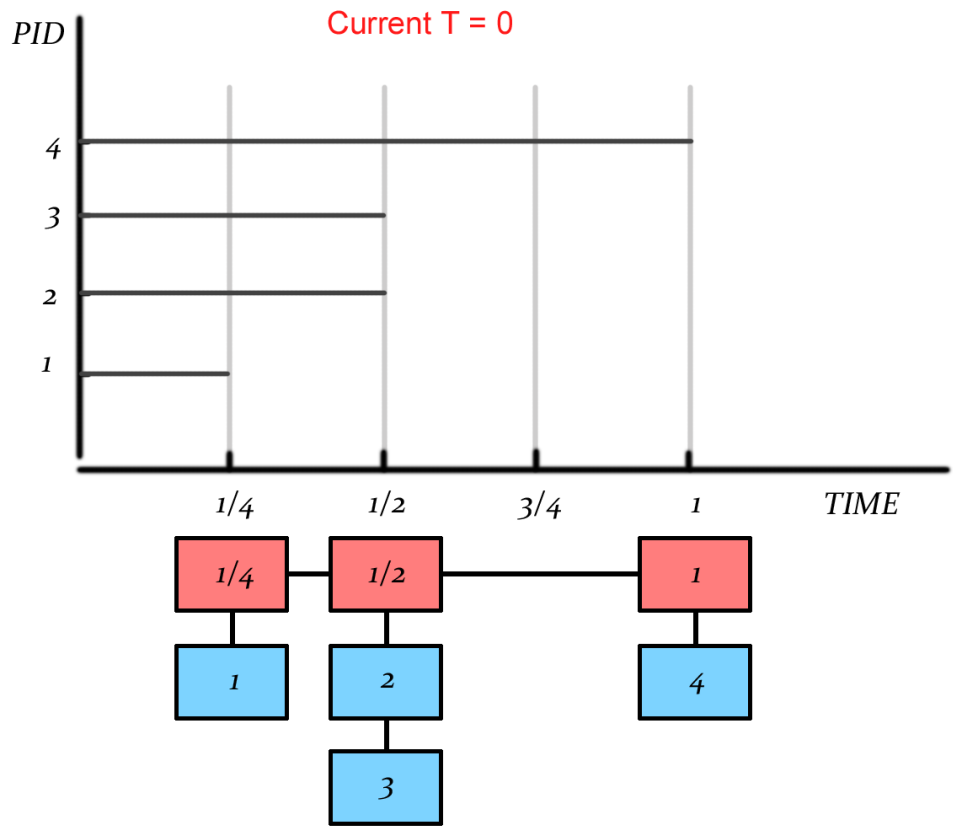}
  \end{center}
  \caption{Sketch of the linked list, together with HITS chart. Horizontal axis is active time $t_i + \Delta t_i$. Vertical axis is particle ID. Red boxes are T-node. Blue boxes are P-node.}
  \label{fig_build}
\end{figure}

At the beginning of simulation, evaluate $\Delta t_i$ for all particles, and set $t_i$ to zero (or checkpoint value) just as normal N-body code does. Construction of the linked list is based on the value of active time of every particle. Particles with same active time are attached to the same T-node, stored in a linked list (P-list). T-nodes with different value of active time are linked in ascending sequence (T-list). Figure \ref{fig_build} gives an example of the linked list. Consider four particles, active time for particle \textbf{1} is $1/4$, for particle \textbf{2} and \textbf{3} are $1/2$, for particle \textbf{4} is $1$. Then we have three T-nodes labeled with $1/4, 1/2$ and $1$, explicitly. Particles are attached to T-nodes according to their active time.

After the construction, a pointer \textbf{*CurrTNode} is pointed (and always pointed) to the first T-node in T-list. This pointer is the only way to access the linked list.

Now we come to the active particle selecting part. As one can see, T-list is already sorted in ascending sequence during the construction procedure. We immediately know the value of \textbf{min\_t} to be the time recorded in the first T-node, and PID for active particles. Here in the example, $\textbf{min\_t} = 1/4$ and only one particle (PID=1) is active in this time step. With these information we continue to the integration part to compute the new position, velocity and $\Delta t$ for active particles. At the end of this time step, set the system time to \textbf{min\_t}.

\begin{figure}[htbp]
  \begin{center}
  \includegraphics[width=0.8\columnwidth]{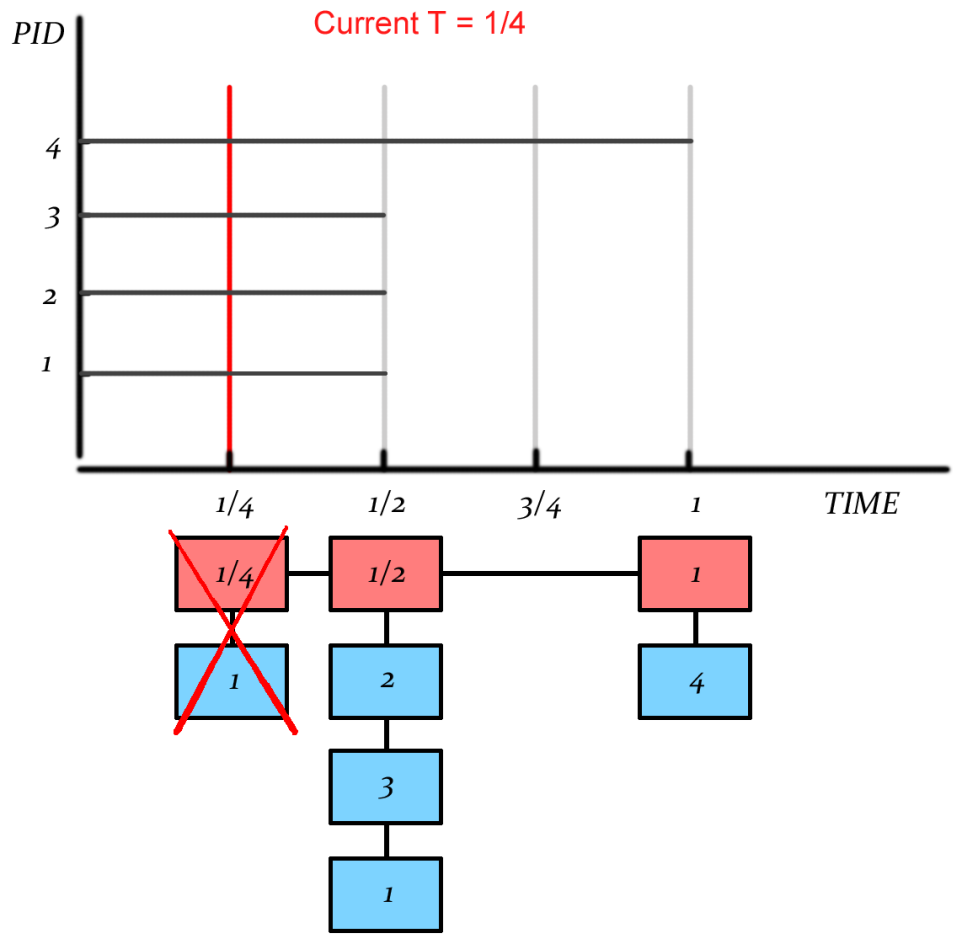}
  \end{center}
  \caption{Sketch of the linked list, together with HITS chart in ``modify" step. Horizontal axis is active time $t_i + \Delta t_i$. Vertical axis is particle ID. Red boxes are T-node. Blue boxes are P-node. Solid red line indicate current system time.}
  \label{fig_select}
\end{figure}

Before advancing to next time step, we modify the linked list according to the new value of active time. As shown in Figure \ref{fig_select}, next active time for particle \textbf{1} is $1/2$. So in the modification step, we attach particle \textbf{1} to the T-node marked with $1/2$ and delete the T-node marked with $1/4$ and its belongings. Also update the value of \textbf{*CurrTNode}.

\begin{figure}[htbp]
  \begin{center}
  \includegraphics[width=0.8\columnwidth]{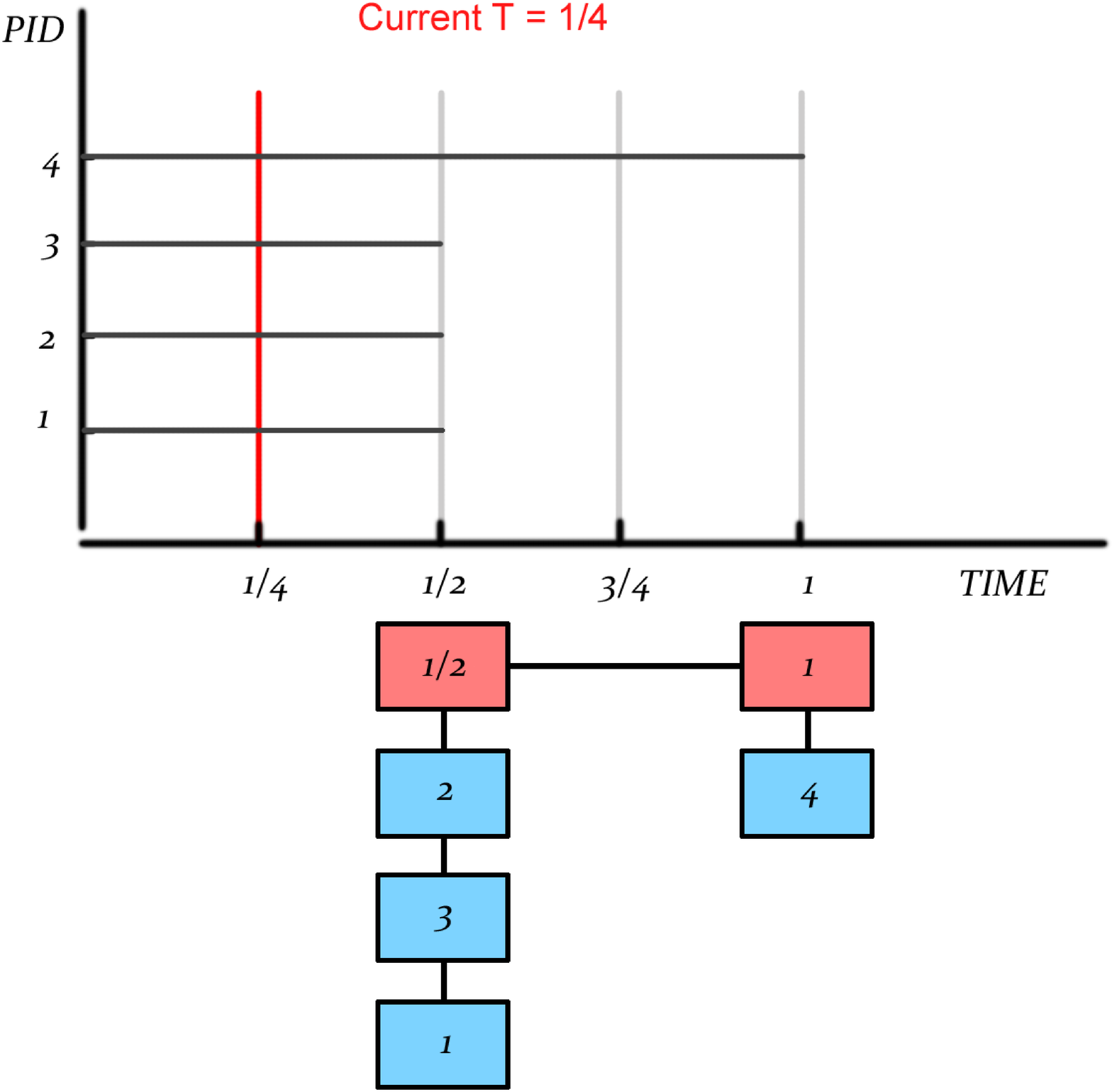}
  \end{center}
  \caption{Sketch of the linked list, together with HITS chart after ``modify" step. Horizontal axis is active time $t_i + \Delta t_i$. Vertical axis is particle ID. Red boxes are T-node. Blue boxes are P-node. Solid red line indicate current system time.}
  \label{fig_next}
\end{figure}

After modification we can enter the next integration loop. Again check the first T-node in T-list and we will know the value of \textbf{min\_t} and PID for the active particles. (Figure \ref{fig_next})

\section{Feasibility of Linklist method}

In this section, we compare the new algorithm with the parallel selection method in both good and bad case. We implement this Linklist algorithm into \PGRAPE code and run with models which are used in scientific studies. In good case, the $N$ particles are distributed following Plummer model with 1 massive particle in the center. Let the initial model evolve, sometimes a small group of particles will become tightly bound to the massive one, greatly reduce $\Delta t$ for every time steps and hence reduce $\overline{N_{act}}$. This situation is referred as bad case. We measure the wall clock time spent for active particle selection within one N-body time integration.

\begin{figure}[htbp]
  \begin{center}
  \includegraphics[width=0.8\columnwidth]{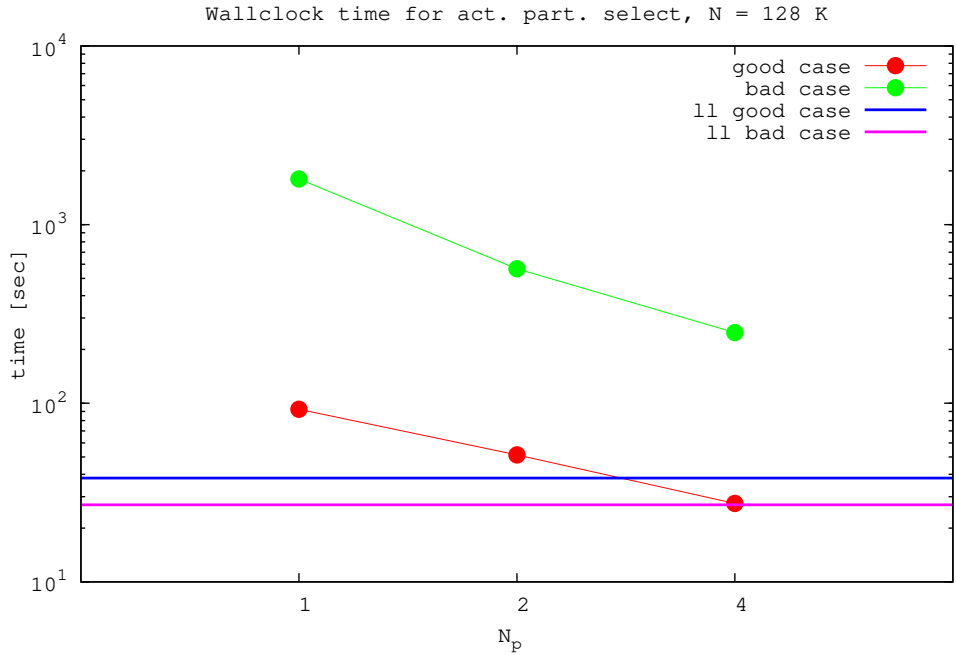}
  \end{center}
  \caption{Wall clock time for active particle selection. Compare between parallel selection (line with symbol) and Linklist method (horizontal lines). Horizontal axis is number of process, vertical axis is execution time in unit of second. The meaning of good case and bad case are explained in the text.}
  \label{fig_comp_128K}
\end{figure}

The Linklist algorithm is performed in serial way, the execution time should be independent of number of process, although in practice it has small variations. For the old method we run it on 1, 2 and 4 nodes (each node has 1 process). The result is shown in Figure \ref{fig_comp_128K}. For 128K particles, $\overline{N_{act}}$ is around 700 in good case, while only $\sim40$ in bad case. One can see the difference between good and bad case for parallel method is almost one order of magnitude for same $N_p$. Because total number of time steps $N_{step}$ becomes roughly 1 order of magnitude higher. And time decrease with $N_p$ as expected. Compare with the results of Linklist method, we see parallel method are slower except one data point where $N_p = 4$. In bad case, parallel method is always much slower. This is mainly caused by the increase of $N_{step}$. $\overline{N_{act}}$ is also small, the data package sent by MPI functions is small which will under use the network and increase the communication cost. For Linklist method, time consumption is lower in bad case than in good case. This is simply because the total number of integrated active particle is lower in bad case ($\sim 5.4\times 10^{7}$, in good case it is $\sim 8.1\times 10^{7}$.) In general, these results set a lower and upper limit for time consumed in active particle selection.

\section{Discussion}

From above description, one can see number of operations is reduced. In the old scheme, number of operations is proportion to $N$. While in this new method, it is proportion to $\overline{N_{act}}$, $\overline{N_{act}} \propto N^{0.613}$ \citep{BNZ2011}. (Note this scaling formula is obtained with Plummer model which is often used for performance test. In practice, one deal with different situations and usually the number of active particle is much smaller.) Thus the new method can reduce the computation complexity for active particle selection, from $O(N\cdot N_{step})$ to $O(\overline{N_{act}}\cdot N_{step})$. And the hit rate is 100\%! We look through $\overline{N_{act}}$ particles (P-list) to select exactly the same number of active particles.

A more recent approach implemented in \PGPU \citep{BNZ2011} sorts the active time array by a fast C++ library call \textbf{std::sort()}. Benefit from this sorting function and the fact that large fraction of the array elements are in right place, the time cost only scale as $O(\overline{N_{act}}\cdot \log(\overline{N_{act}}))$ instead of $O(N\cdot \log(N))$. After sorting, the whole array contains many segments, particles inside each segment have same active time. The first segment of the array stores the active particle's information. Time cost for reading these elements scales as $N_{act}$. Compare to sorting method, Linklist method is roughly one order of magnitude slower, although the computation cost scaling is similar. We note that in \PGPU, sorting is done within an array which has good locality and can fully use CPU cache to accelerate execution. While locality of linked list may be worse than array and will cause more latency in CPU execution.


\bibliographystyle{apj}
\bibliography{act_def_ll}

\end{document}